\documentclass[aip,twocolumn,reprint]{revtex4-1}
\usepackage{color}
\usepackage{graphicx}
\usepackage{amsmath}
\usepackage{amssymb}

\renewcommand{\vec}[1]{\mathbf{#1}}

\newcommand{\ddt}[2]{\frac{\partial{#1}}{\partial{#2}}}

\newcommand{\comment}[1]{}
\newcommand{\ig}[2]{\includegraphics[width = #1]{#2}}

\begin{document}
\title{The island coalescence problem: scaling of reconnection in extended fluid models including higher-order moments}
\author{Jonathan Ng}
\author{Yi-Min Huang}
\author{Ammar Hakim} 
\author{A. Bhattacharjee}
\affiliation{Center for Heliophysics, Princeton Plasma Physics Laboratory, Princeton, NJ 08540, USA}
\author{Adam Stanier}
\author{William Daughton}
\affiliation{Los Alamos National Laboratory, Los Alamos, New
Mexico 87545, USA}
\author{Liang Wang}
\author{Kai Germaschewski}
\affiliation{Space Science Center and Physics Department, University of New Hampshire, Durham, New Hampshire 03824, USA}

\date{\today}
\begin{abstract}
As modeling of collisionless magnetic reconnection in most space plasmas with realistic parameters is beyond the capability of today's simulations, due to the separation between global and kinetic length scales, it is important to establish scaling relations in model problems so as to extrapolate to realistic scales. Recently, large scale particle-in-cell (PIC) simulations of island coalescence have shown that the time averaged reconnection rate decreases with system size, while fluid systems at such large scales in the Hall regime have not been studied. Here we perform the complementary resistive MHD, Hall MHD and two fluid simulations using a ten-moment model with the same geometry. In contrast to the standard Harris sheet reconnection problem, Hall MHD is insufficient to capture the physics of the reconnection region. Additionally, motivated by the results of a recent set of hybrid simulations which show the importance of ion kinetics in this geometry, we evaluate the efficacy of the ten-moment model in reproducing such results.
\end{abstract}
\maketitle

\section{Introduction}
Magnetic reconnection is a change in topology of the magnetic field lines in a plasma \cite{dungey:1953}, often with the conversion of stored magnetic energy to the kinetic energy of accelerated particles. It is believed to play an important role in many laboratory and astrophysical plasma processes, including sawtooth crashes in tokamaks, solar flares, magnetic substorms in the Earth's magnetosphere and coronal mass ejections \cite{goeler:1974,vasyliunas:1975,phan:2000,bhattacharjee:2001,yamada:2010}. 

For reconnection to take place, the motion of the plasma must decouple from the magnetic field lines. In collisionless environments such as the magnetosphere, this takes place in the electron diffusion region, and due to the kinetic scales involved, cannot be described purely by resistive magnetohydrodynamic (MHD) models such as the Sweet-Parker model \cite{sweet:1958, parker:1957}. As large-scale particle-in-cell (PIC) simulations of collisionless reconnection are computationally prohibitive due to the difficulties in resolving the kinetic length-scales, it is important to establish scaling relations in model problems so as to extrapolate to realistic systems. 

One such geometry is that of the coalescence problem, in which two magnetic islands approach each other and merge \cite{finn:1977,pritchett:1979,biskamp:1980,biskamp:1982}. Although this problem has been studied in the past using both fluid and kinetic codes \cite{bhattacharjee:1983,mandt:1994,dorelli:2001,dorelli:2003b,knoll:2006,knoll:2006prl, pritchett:2007,simakov:2009}, a detailed comparison between the models was not performed due to the different scales of these studies. In particular, Hall MHD studies have not been performed for extremely large systems. 

A recent set of kinetic simulations has highlighted the effect of system size on the coalescence process and the differences between fluid and kinetic models \cite{karimabadi:2011}. In the kinetic simulations, the peak reconnection rate is independent of island size for larger systems, while the average reconnection rate (over two global Alfv\'en times from the start of the simulation) shows a $\sqrt{d_i/\lambda}$ scaling \cite{karimabadi:2011}, where $d_i$ is the ion inertial length and $\lambda$ is proportional to the length scale of the islands. On the other hand, while Hall MHD simulations show a decrease in the reconnection rate for larger systems, they do not remain in the regime of Hall reconnection as system size is increased \cite{dorelli:2003b,knoll:2006}. Theoretical scalings have been derived for driven reconnection in the Hall MHD regime for simple geometries \cite{wang:2001,dorelli:2003}, but it is not clear that they are relevant to island coalescence. 

In view of the PIC results on the coalescence instability, it is timely to compare these results with those from resistive MHD, Hall MHD, and higher-order moment equations which attempt to incorporate kinetic effects \cite{yin:2001,wang:2015}. The results of such comparisons provide a basis for evaluating the efficacy of extended fluid models in including kinetic effects in global simulations. In this paper we compare resistive MHD, Hall MHD and ten-moment fluid simulations of the coalescence problem to kinetic simulations, finding discrepancies between the fluid and PIC results that have implications for the efficiency of merging and reconnection in large systems. 

Our results show that in the Hall MHD limit, the current sheet shrinks to an x-point geometry and the average and maximum reconnection rates depend only weakly on the system size, in contrast to the PIC simulations. In the two-fluid simulations, which include both the ion and electron pressure tensors, the current sheets are wider and approach the island size. Additionally, motivated by the results of new kinetic and hybrid simulations \cite{stanier:2015} which demonstrate the importance of ion kinetics in this geometry, we show that allowing the ion pressure tensor to deviate from isotropy in the ion diffusion region improves the agreement between the two-fluid and PIC results for small systems. However, the eventual formation of secondary islands in larger systems causes reconnection to proceed to completion and a weak dependence of reconnection rate on system size. 

The remainder of this paper is organised as follows: In Section~\ref{sec:coal} we describe the coalescence geometry and parameters used in our comparison between the fluid and PIC simulations. The resistive and Hall MHD models and the results of these simulations are discussed in Section~\ref{sec:mhd}. We then describe the ten-moment two-fluid model and simulations in Section~\ref{sec:gke}, after which our results are summarised.

\section{Island coalescence}
\label{sec:coal}

We perform resistive MHD, Hall MHD and two-fluid simulations of the coalescence problem in order to compare the system size scaling of the reconnection rate with the PIC simulation results.

The simulations are translationally symmetric in the $y$ direction, and the initial configuration is a Fadeev equilibrium \cite{fadeev:1965}, described by the vector potential and density
\begin{equation}
\begin{split}
A_y &= -\lambda B_0 \ln \left[\cosh(z/\lambda) + \epsilon \cos(x/\lambda)\right]\\
n &= n_0 (1-\epsilon^2)/\left[\cosh(z/\lambda) + \epsilon \cos(x/\lambda)\right]^2 + n_b.
\end{split}
\end{equation}

Here $B_0$ is the $x$-component of the magnetic field upstream of the layer, $\epsilon$ controls the island size and $\lambda$ is the half width of the current sheet. We use the same physical parameters as described in Ref.~\onlinecite{karimabadi:2011}, with $\epsilon = 0.4$, which corresponds to an island half-width of approximately $1.2\lambda$, and background density $n_b = 0.2 n_0$. The system size is $L_x \times L_z = 4\pi\lambda \times 2\pi\lambda$, with periodic boundary conditions in the $x$ direction. Conducting walls for fields and reflecting walls for particles are used in the $z$ direction. 

Where applicable in the fluid models, we use mass ratio $m_i/m_e = 25$, electron thermal speed $v_{the}/c = 0.35$ and $T_i = T_e = T$, and the value of $T$ is set by the upstream equilibrium condition $\beta = 1$. The ratio of electron plasma frequency to gyrofrequency is $\omega_{pe}/\Omega_{ce} = 2$. The system is unstable to the ideal coalescence instability \cite{finn:1977}, and reconnection is initiated by a perturbation with amplitude $0.1 B_0$ \cite{daughton:2009,karimabadi:2011}.

Simulations are run for $2.5 t_A$, where $t_A = L_x/V_A$ is the global Alfv\'en time, where $V_A$ is calculated using $n_0$ and $B_0$. In the same manner as Ref.~\onlinecite{karimabadi:2011}, the normalised reconnection rate is determined by $E_R = (1/B' V_A')\partial\psi/\partial t$ where $B'$ and $V_A'$ are calculated using the maximum magnetic field between the centres of the two islands at $t = 0$. The flux within an island is defined as the difference between $A_y$ at the X- and O-points $\psi \equiv [\text{max} A_y]_{X} -  [A_y]_{O}$. The average reconnection rate is calculated during the interval between $t = 0$ and $t = 1.5 t_A$.

The PIC results reveal that after an initial decrease, the maximum reconnection rate remains approximately constant in large systems while the average reconnection rate shows a decrease with system size. The reconnection rates used for comparison in this paper are from Ref.~\onlinecite{stanier:2015}, in which the average rate scales like $(d_i/\lambda)^{0.8}$. This differs from that described in Ref.~\onlinecite{karimabadi:2011} due to the influence of the aspect ratio of the simulation box \cite{stanier:2015}. In all the simulations here the box aspect ratio is kept constant as the system size is increased.

\section{MHD Simulations}
\label{sec:mhd}

The resistive and Hall MHD models used in our simulations are described by the following system of equations normalised to characteristic values of $n_0, B_0$:
\begin{equation}
\begin{split}
\partial_t\rho + \nabla\cdot\left(\rho \vec{v}\right) &= 0\\
\partial_t(\rho\vec{v}) + \nabla\cdot(\rho \vec{vv}) &= \vec{J}\times\vec{B} -\nabla P +\mu\nabla^2(\rho\vec{v}) \\
\vec{E} + \vec{v}\times\vec{B} &= \frac{d_i}{\rho} (\vec{J}\times\vec{B} - \nabla P_e) + \eta\vec{J} - \eta_H \nabla^2\vec{J}\\
\nabla\times\vec{E} &= -\partial_t\vec{B}.
\end{split}
\label{eq:hmhd}
\end{equation}
Here $P = P_e + P_i = 2\rho T$ is the scalar pressure, and we assume an adiabatic equation of state with $\gamma = \tfrac{5}{3}$. $d_i = 1$ is the ion inertial length (except in the purely resistive simulations), $\mu = 10^{-2}$ is the normalised ion viscosity, $\eta$ is the resistivity and $\eta_H$ is the hyper-resistivity. Spatial derivatives are approximated using a five-point stencil, while time stepping uses a trapezoidal leapfrog method. A small numerical noise is added to the initial ion momentum in order to observe the formation of secondary islands \cite{huang:2010}. 

In earlier Hall MHD studies of coalescence \cite{dorelli:2003b,mandt:1994}, the merging process undergoes a transition from Hall to resistive as system size is increased. Ref.~\onlinecite{dorelli:2003b} observes that at a constant Lundquist number, for small systems in the electron-MHD regime with $d_i \gtrsim \lambda \gg \delta$, where $\delta$ is the current sheet width, the maximum reconnection rate scales like $\lambda^{-7/3}$, while for larger $\lambda/d_i$, there was a transition to the resistive regime ($\lambda \gg \delta \gtrsim d_i$), with slower reconnection rates \cite{dorelli:2003b}. In Ref.~\onlinecite{mandt:1994} a strong scaling of $\lambda^{-2}$ in the electron-MHD regime is observed, though the result may have been influenced by resistive diffusion \cite{pritchett:2007, dorelli:2003b}. Theoretical scalings for driven reconnection using the Hall MHD model in simple current sheet geometries predict a rate that scales as $\sqrt{d_i/\lambda}$ \cite{wang:2001, dorelli:2003}, but, in this paper, we find that this scaling does not agree with simulation results for the island coalescence problem. 

The regime expected to be most relevant when performing the comparison with PIC simulations is the limit with $\delta \ll d_i \ll \lambda$. Earlier studies \cite{dorelli:2003b,simakov:2009} have shown the importance of having $\delta \ll d_i$ for reconnection rates independent of dissipation in the coalescence problem, but a system size scaling for large systems while maintaining this condition has not been done. We present results where $\delta$ is set by resistivity, and results from two different simulation codes where $\delta$ is set by hyper-resistivity \cite{stanier:2015}.

\subsection{Resistive MHD}
We first present the results of the purely resistive MHD simulations (setting $d_i =0, \eta_H =0$ in Ohm's law). While a direct comparison between this model and the PIC simulation is not appropriate due to the numerous approximations made, it is useful as a comparison to the Hall MHD models which are studied below. This system has been studied using reduced MHD \cite{knoll:2006}, in which flux pileup -- the buildup of magnetic energy just upstream of the current sheet -- was observed at small resistivities. This caused sloshing, or bouncing, of the islands and periodic stalling of the coalescence process. In other MHD studies of coalescence, the reconnection current sheet became unstable to the tearing instability \cite{biskamp:1982}, causing the formation of secondary magnetic islands, or plasmoids, within the current sheet.

\begin{figure}[t]
\ig{3.375in}{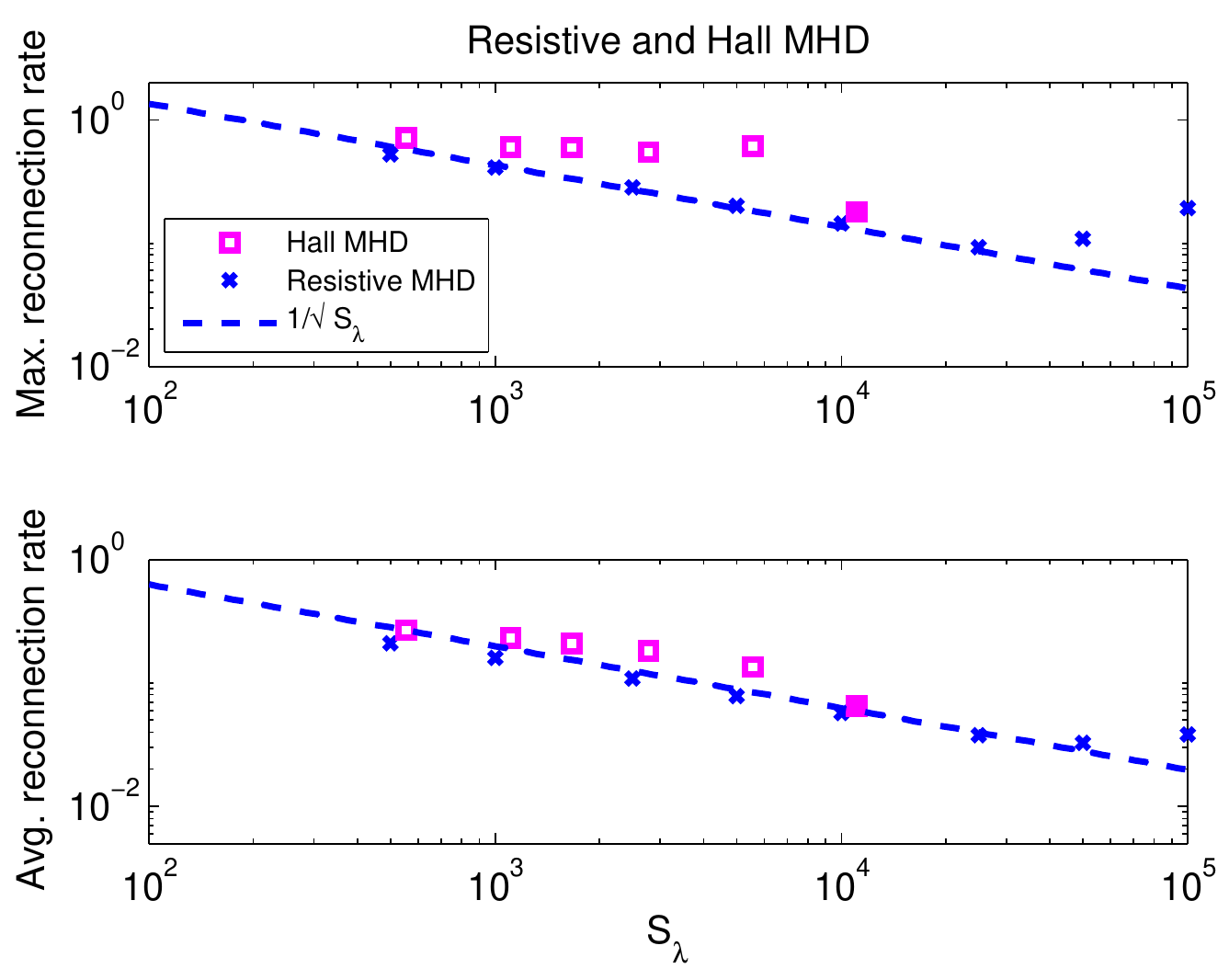}
\caption{Comparison of maximum and average reconnection rates in resistive and resistive Hall MHD. The filled square shows the case where the current sheet does not thin sufficiently to allow Hall reconnection to take place. The dashed line shows the Sweet-Parker scaling. In the resistive case, plasmoids are observed for $S_\lambda \geq 5\times10^4$.}
\label{fig:resistive}
\end{figure}

Our results are shown in Fig.~\ref{fig:resistive}, and we observe a similar Sweet-Parker like $S_\lambda^{-1/2}$ scaling of the maximum reconnection rate. Here the definition of the Lundquist number uses $\lambda$, which is a proxy for the length of the current sheet, as the length scale in $S_\lambda \equiv \mu_0V_A\lambda/\eta$, while $V_A$ is calculated using $n_0, B_0$. In the resistive simulations the resistivity and viscosity are kept constant as $\lambda$ increases. For larger systems with $S_\lambda \geq 5\times 10^4$, we observe the onset of the plasmoid instability, similar to the results of Ref.~\onlinecite{biskamp:1982}, causing the maximum reconnection rate to be enhanced compared to the Sweet-Parker scaling \cite{huang:2010}.

\subsection{Hall MHD}
Separation of ion and electron flow is present in Hall MHD, allowing fast reconnection to take place \cite{birn:2001}. Earlier studies using the coalescence geometry have shown that if the current sheet width falls below the ion skin depth, Hall physics becomes important and reconnection rates are faster than in the resistive limit \cite{knoll:2006prl, dorelli:2001, dorelli:2003b}. 

\begin{figure}[t]
\ig{3.375in}{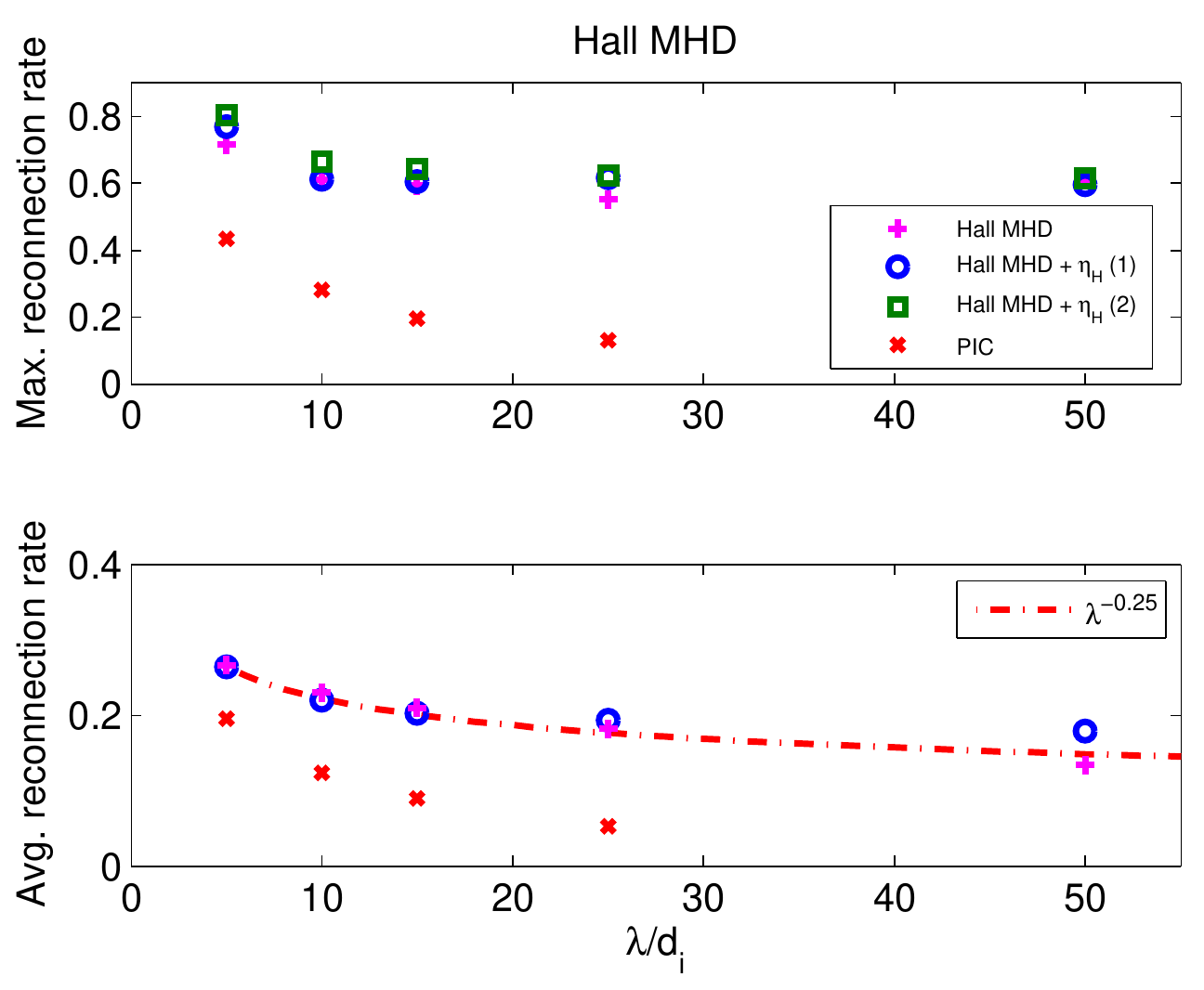}
\caption{Comparison of maximum and average reconnection rates between PIC and resistive Hall MHD models. The Hall MHD results with hyper-resistivity are from the codes described in (1) this paper and (2) Ref.~\onlinecite{stanier:2015}. PIC data are from Ref.~\onlinecite{stanier:2015}. }
\label{fig:resistiveHall}
\end{figure}

The comparison between the resistive Hall MHD results and purely resistive MHD results is shown in Fig.~\ref{fig:resistive}, while the comparison to the PIC results is shown in Fig.~\ref{fig:resistiveHall}. We keep the resistivity constant at $\eta = 9\times 10^{-3}$ as $\lambda/d_i$ is increased. The results with hyper-resistivity are discussed later in this section. 

In the simulations with $\lambda/d_i \leq 25$, the current sheet quickly shrinks to an x-point, with $\delta < 0.1 d_i$, and other than the $\lambda = 5 d_i$ simulation which shows a larger reconnection rate, the aspect ratio at the time of maximum reconnection rate remains constant $\delta/l \approx 0.07$. Aside from a relatively large initial decrease in maximum reconnection rate from $\lambda/d_i = 5$ to $\lambda/d_i = 10$, the rate only decreases slightly as system size is increased when compared to the purely resistive simulations. The average reconnection rate decreases like $\sim (d_i/\lambda)^{0.25}$, which is much weaker than the PIC scaling \cite{stanier:2015, karimabadi:2011}. It should be noted that the scaling observed here is different from that in Ref.~\onlinecite{dorelli:2003b} as our current layer remains much thinner than the ion inertial length for $\lambda/d_i \leq 50$, while theirs undergoes a transition to the resistive regime. 

At $\lambda/d_i = 50$, the system initially behaves similarly to resistive MHD, with a Sweet-Parker current sheet on the $d_i$ scale forming initially. At approximately $t = 1.1 t_A$, the current sheet rapidly shrinks to an x-point, and the maximum reconnection rate becomes comparable to that observed in the smaller systems. Using the same value of $\eta$, at $\lambda/d_i = 100$, the current sheet does not thin sufficiently to allow Hall reconnection to take place, which causes both average and maximum reconnection rates to be closer to the resistive MHD rates. As such, it is only shown in Fig.~\ref{fig:resistive}. Sloshing is observed in the largest Hall MHD simulations ($\lambda/d_i \geq 50$), which may be due to insufficient separation of $d_i$ and $\delta$ and larger flux pileup \cite{knoll:2006prl}. In contrast, sloshing is observed for $\lambda/d_i \geq 10$ in the PIC simulations even with sufficient separation of ion and electron scales as the reconnection rate is too small for the islands to coalesce fully initially \cite{stanier:2015}. In both cases the sloshing begins slightly after $t\approx t_A$, close to the time of maximum flux pileup \cite{karimabadi:2011,stanier:2015}. 

\begin{figure}
\ig{3.375in}{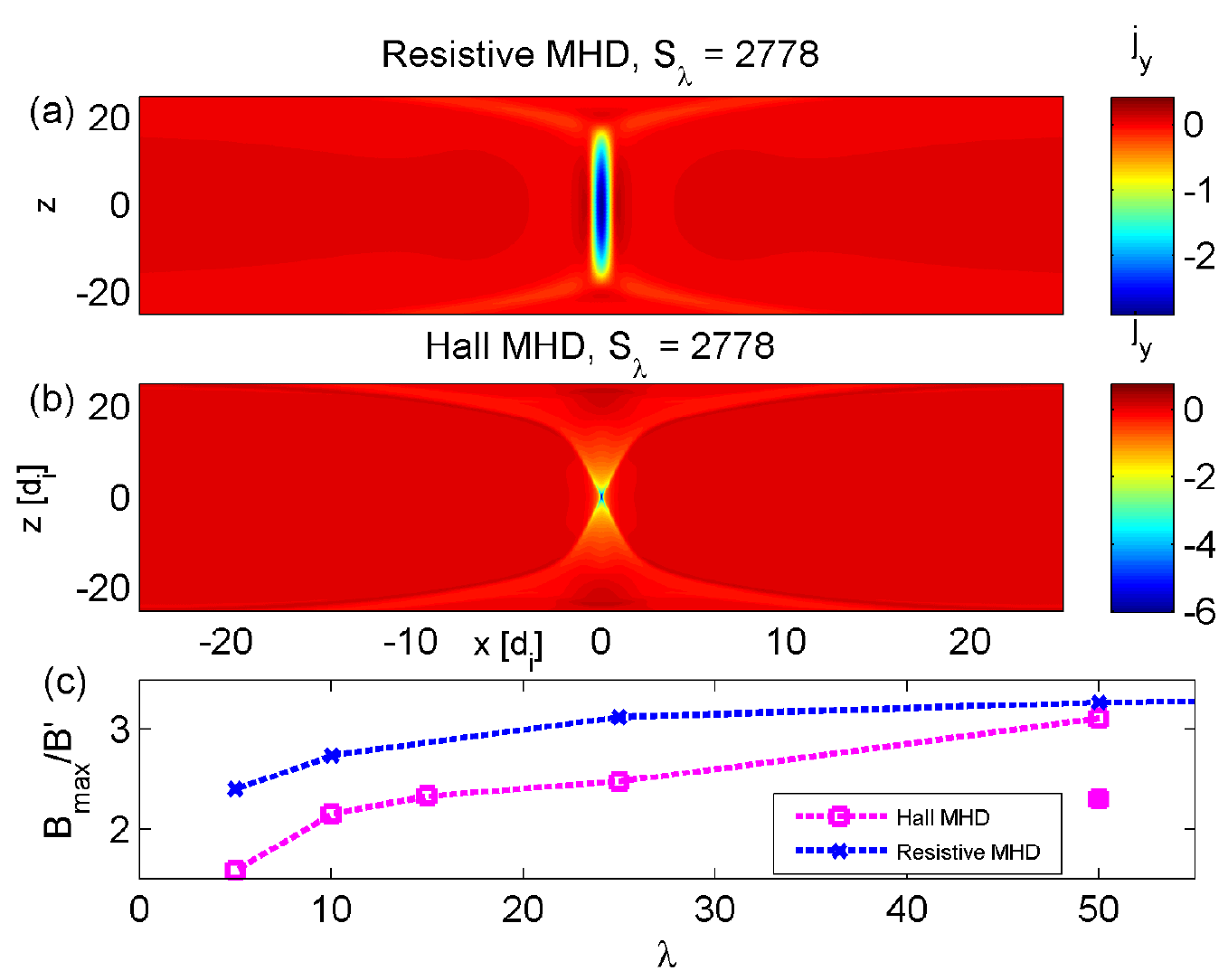}
\caption{Out-of-plane current density at $t = 0.99 t_A$. (a) Resistive MHD with $S_\lambda = 2778$. As there is no $d_i$ here, units are normalised for direct comparison with the $\lambda/d_i = 25$ resistive Hall simulation. (b) Resistive Hall MHD with $\lambda/d_i = 25$, $\eta = 9 \times 10^{-3}$ and $S_\lambda = 2778$. (c) Comparison of maximum flux pileup between resistive and Hall MHD simulations. The filled square for $\lambda = 50 d_i$ is the pileup after the transition to Hall reconnection at the time of maximum reconnection rate. }
\label{fig:mhdcurrent}
\end{figure}

The structure of the current sheet at $t = 0.99 t_A$, when the reconnection rate is close to its maximum, is shown in Fig.~\ref{fig:mhdcurrent}. The top panel shows the Sweet-Parker current sheet formed in the purely resistive simulation. As expected, its length approaches the island size and the aspect ratio shows the typical $1/\sqrt{S_\lambda}$ scaling. The localised current sheet characteristic of Hall reconnection is shown in Fig.~\ref{fig:mhdcurrent}(b). As mentioned earlier, the current sheet width is much less than $d_i$ and the maximum reconnection rate shows only a weak scaling with system size for small systems.

Fig.~\ref{fig:mhdcurrent}(c) shows the differences in flux pileup between the resistive and Hall MHD simulations. For Hall MHD the pileup occurs on the edge of the ion layer, and below this scale it is reduced as the field is frozen in to the faster electron flow \cite{dorelli:2003, knoll:2006prl}. This results in lower peak values compared with resistive MHD. In both models, the pileup saturates for larger systems.

Finally, the results of the Hall MHD simulations with hyper-resistivity are shown in Fig.~\ref{fig:resistiveHall}. In this set of simulations, $\eta_H = 10^{-4}$ was kept constant, while the value of resistivity was small but finite at $\eta = 10^{-5}$. A scan in $\eta$ showed that the reconnection rate and current sheet geometry is insensitive to its exact value as the hyper-resistive term dominates. In these simulations, the scale of the current sheet $\delta \approx 0.2 d_i$ is sufficiently separated from the ion inertial length and Hall reconnection is observed for the whole range of system sizes.

%\begin{figure}[ht]
%\ig{3.375in}{hyperHall}
%\caption{Scaling of maximum and average reconnection rate in Hall MHD with hyper-resistivity as compared to PIC simulations. The Hall MHD results are from the codes described in (1) this paper and (2) Ref.~\onlinecite{stanier:2015}.}
%\label{fig:hyperHall}
%\end{figure}

As shown in Fig.~\ref{fig:resistiveHall}, the maximum reconnection rate remains approximately constant as system size is increased for $\lambda \geq 10 d_i$, and the average rate scales like $(d_i/\lambda)^{0.2}$, which is consistent with our earlier resistive Hall MHD results. These results have been compared with a second Hall MHD code \cite{stanier:2015}, which solves a slightly different set of equations, in which the ion viscosity term in the momentum equation is replaced by one proportional to $\nabla^2\vec{v}$ and an energy equation with heating and conduction is used. Despite these differences, the maximum reconnection rates are in good agreement.

\subsection{MHD summary}
The MHD simulations show the different regimes of coalescence in large systems. In the resistive limit, flux pileup outside the current sheet causes sloshing, and the maximum reconnection rate shows an approximate $S_\lambda^{-1/2}$ scaling. For $S_\lambda < 5\times 10^4$, the results are in agreement with previous work on the coalescence problem \cite{knoll:2006}, while for larger systems we observe the eventual formation of secondary islands \cite{biskamp:1982}.

Adding the Hall term to the generalised Ohm's law reduces the pileup \cite{dorelli:2003} and increases the reconnection rate. When the resistivity is sufficiently small so that the current sheet thickness is much smaller than the ion skin depth, an x-point geometry is observed and both maximum and average reconnection rates decrease weakly with the system size. Similar conclusions are reached in the hyper-resistive simulations, which reinforces the idea that the rate is set by the ion scale physics and does not depend on the mechanism by which the frozen-in condition for electrons is broken. 

These results are in contrast to the PIC and hybrid (kinetic ions, fluid electrons) simulations \cite{karimabadi:2011,stanier:2015}, in which the maximum reconnection rate is independent of system size for large systems, but the average rate decreases with size. Quantitatively, the magnitudes of the reconnection rates observed in both resistive and hyper-resistive Hall MHD models are larger than in the PIC simulations. There are also differences in the geometry of the current sheets. In the Hall dominated regime we observe the formation of localised x-points, which differs from the PIC simulations, where the current sheet length approaches $\lambda$ \cite{karimabadi:2011}. 

\section{Two-fluid simulations using the ten-moment model}
\label{sec:gke}

In a fully kinetic treatment, the electron pressure is generically a tensor with non-vanishing off-diagonal elements. It has recently been shown that electron pressure anisotropy is important for setting the structure of the reconnection current sheet in elongated Harris-like current sheets \cite{egedal:2013,le:2013,ohia:2012}, while the off-diagonal elements balance the reconnection electric field in the generalised Ohm's law \cite{yin:2001,swisdak:2005}. 

To include the effect of the pressure tensor, a ten-moment model has been developed by Wang et al.~\cite{wang:2015}, and implemented using the numerical code Gkeyll \cite{hakim:2006,hakim:2008}. In Ref.~\onlinecite{wang:2015} it has been demonstrated for Harris sheet initial conditions that a number of features of the electron diffusion region, such as the thickness and outflow velocity, from the fully kinetic results can be approximated by the ten-moment model. Here we describe how well the model performs for the island coalescence geometry.

For each species, the fluid equations are obtained by taking moments of the Vlasov equation. The continuity and momentum equations are as follows:
\begin{equation}
\begin{split}
\ddt{n}{t} + \ddt{}{x_j}(nu_j) &= 0\\
m \ddt{}{t}(nu_i) + \ddt{\mathcal{P}_{ij}}{x_j} &= nq(E_j + \epsilon_{ijk}u_jB_k). \\
\end{split}
\end{equation}
where $\mathcal{P}_{ij}$ is the second moment of the distribution function
\begin{equation}
\mathcal{P}_{ij} \equiv m\int v_iv_j f d^3v
\end{equation}

The ten moment model evolves the full pressure tensor according to
\begin{equation}
\ddt{\mathcal{P}_{ij}}{t} + \ddt{\mathcal{Q}_{ijk}}{x_k} = nqu_{[i}E_{j]} + \frac{q}{m}\epsilon_{[ikl}\mathcal{P}_{kj]}B_l,
\end{equation}
where $\mathcal{Q}_{ijk}$ is the third moment of the distribution function 
\begin{equation}
\mathcal{Q}_{ijk} \equiv m\int v_iv_jv_k f d^3v,
\end{equation}
and the square brackets denote a sum over permutations of the indices (e.g.~$u_{[i}E_{j]} = u_iE_j + u_jE_i$). Following Ref.~\onlinecite{wang:2015} one can write $\mathcal{Q}_{ijk}$ in terms of the heat flux tensor $Q_{ijk} \equiv m\int (v_i-u_i)(v_j-u_j)(v_k-u_k) f d^3v$
\begin{equation}
\mathcal{Q}_{ijk} = Q_{ijk} + u_{[i}\mathcal{P}_{jk]} - 2 m n u_iu_ju_k,
\end{equation}
with the heat flux defined by 
\begin{equation}
\partial_mQ_{ijm} = v_t|k|(P_{ij}-p\delta_{ij}),
\label{eq:heatflux}
\end{equation}
where $P_{ij}$ is the pressure tensor and $p = \text{trace}(P_{ij})/3$ is the scalar pressure. This relaxes the pressure tensor to an isotropic pressure at a rate given by $|k|v_t$, effectively allowing deviations from isotropy at length scales less than $1/|k|$. These equations coupled to Maxwell's equations describe the time evolution of the plasma. 

The model for heat flux used here is similar to that used in earlier studies of reconnection \cite{hesse:1995,yin:2001} and it can be extended to include kinetic effects \cite{hammett:1990}. Proper evaluation of the heat flux requires non-local integration \cite{hammett:1990} or some approximation thereof \cite{dimits:2014}, and further work on this aspect will be necessary. 

\begin{figure}
\ig{3.375in}{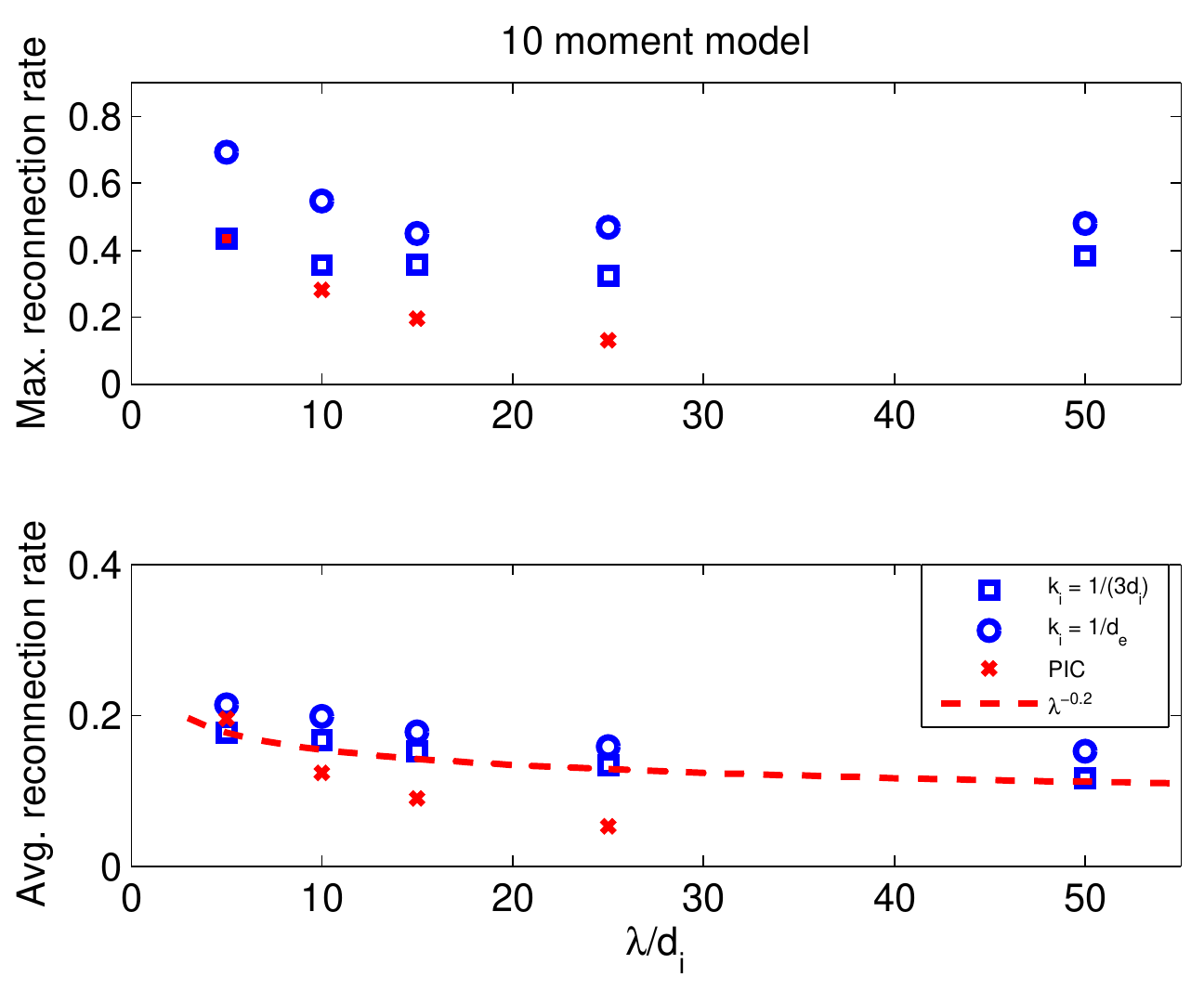}
\caption{Maximum and average reconnection rates in the two-fluid, ten-moment model. }
\label{fig:gke10}
\end{figure}

\begin{figure}
\ig{3.375in}{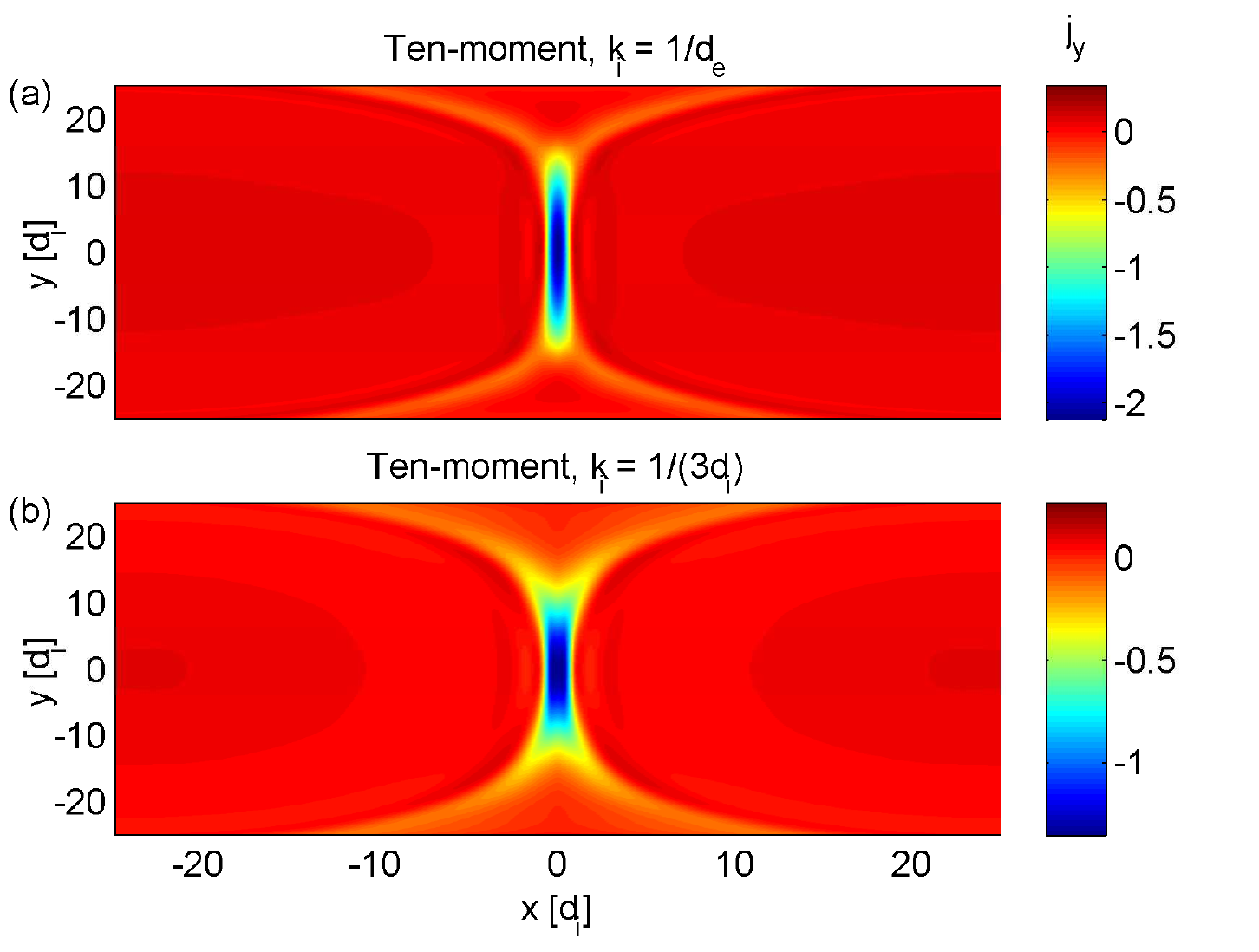}
\caption{Out-of-plane current density in the ten-moment simulations for $\lambda/d_i = 25$ at $t = t_A$ using (a) $k_i = 1/d_e$ (b) $k_i = 1/(3d_i)$. }
\label{fig:gkecurrent_final}
\end{figure}

The results of the ten-moment simulations are shown in Figs~\ref{fig:gkecurrent_final} and~\ref{fig:gke10}. With $k_{e,i} = 1/d_e$, which was used in the Harris sheet study \cite{wang:2015}, the maximum reconnection rates decrease up to $\lambda = 25 d_i$ before levelling off, which is in part due to the formation of secondary islands, which appear for $\lambda \geq 15 d_i$ in the fluid simulations but are not observed in the PIC simulations. There is an approximate factor of two difference between the PIC and fluid reconnection rates. Motivated by the results of a hybrid study of island coalescence \cite{stanier:2015}, a second set of fluid simulations using $k_i = 1/(3d_i)$ was performed.  This allows the pressure tensor to deviate from isotropy at ion scales, which allows a better comparison with hybrid and PIC results. As shown in Fig.~\ref{fig:gke10}, this gives better agreement with the PIC results for small systems. However, for larger systems we observe the formation of secondary islands, and the scaling of the average reconnection rate is much weaker ($\propto (d_i/\lambda)^{0.2}$). 

The structure of the current sheets at $t = t_A$ for both values of $k_i$ is shown in Fig.~\ref{fig:gkecurrent_final}, and compared to the Hall MHD current sheet for the same system size in Fig.~\ref{fig:mhdcurrent}(b), they are broader and more elongated, and consequently have a lower peak current density. There are slight differences between Figs.~\ref{fig:gkecurrent_final}(a) and (b) due to the different $k_i$, which will be discussed in the next section. 

Unlike the resistive MHD and PIC simulations, there is little to no bouncing or stagnation after the reconnection rate reaches its initial maximum, even when the secondary islands form. Instead, they are ejected rapidly, followed by the formation and ejection of more islands, which allows reconnection to continue and proceed to completion, causing larger average reconnection rates to be observed.

\subsection{Role of the ion pressure tensor}

In addition to the difference in the reconnection rates, it is shown in Ref.~\onlinecite{stanier:2015} that kinetic ion effects influence the structure of the ion diffusion region via gradients in the off-diagonal elements of the ion pressure tensor. In particular, the  width of the diffusion region was observed to be 2-3 $d_i$ when $T_i = T_e$ and determined by the unmagnetised ion orbits. 

Here we study how the choice of $k_i$ influences the ion diffusion region geometry by modifying the value of $k$ in Eq.~\eqref{eq:heatflux}, the inverse of the length scale below which the pressure tensor is allowed to depart from isotropy. In the Harris sheet problem, the choice of $k_e = 1/d_e$ captures the physics in the electron diffusion region and reproduces the effect of the electron pressure tensor in balancing the reconnection electric field \cite{wang:2015}. In the coalescence geometry, we find that the reconnection rate is more sensitive to the choice of $k_i$ than $k_e$.

\begin{figure}
\ig{3.375in}{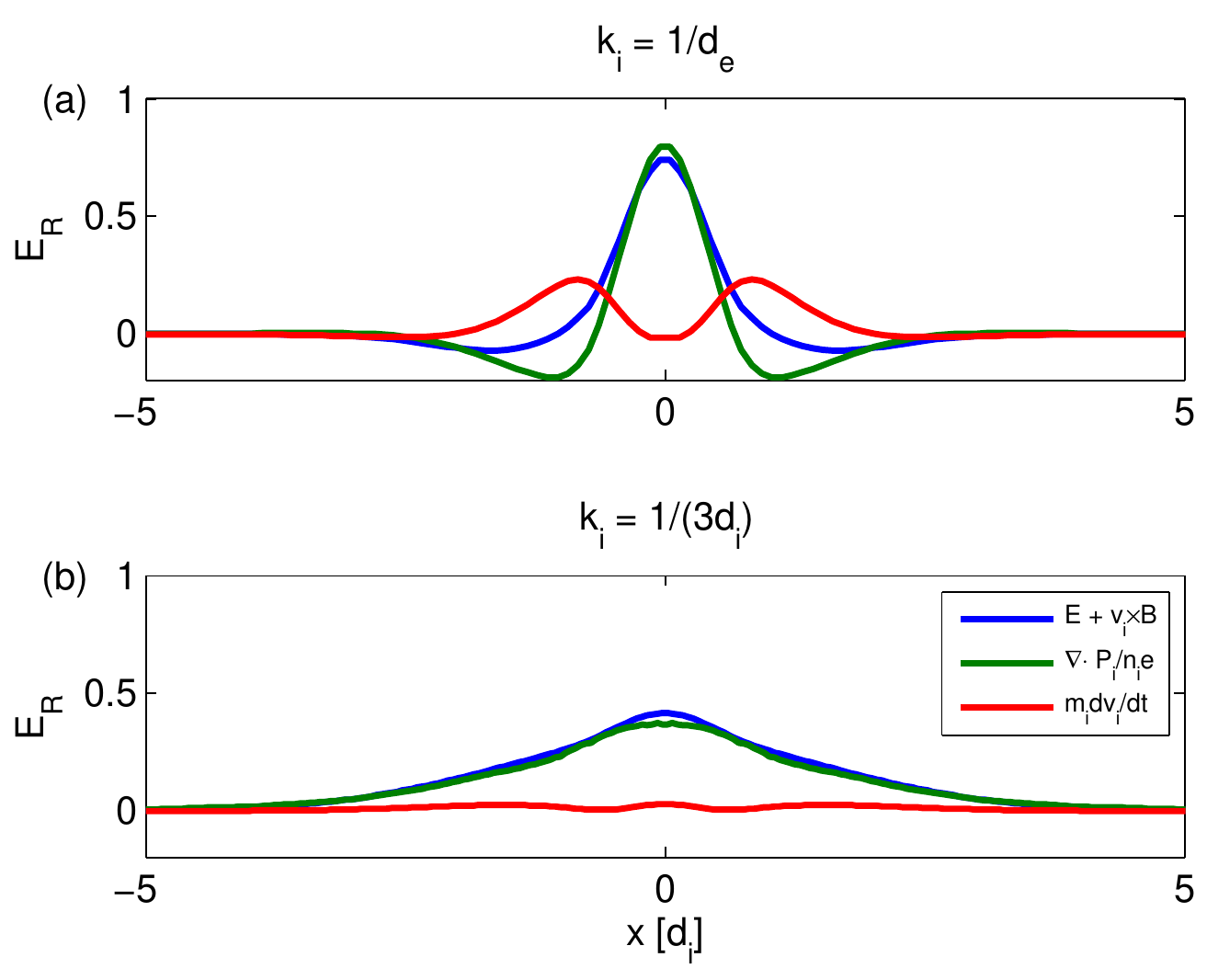}
\caption{Effect of $k_i$ on the ion momentum equation. Reducing $k_i$ from (a) $1/d_e$ to (b) $1/(3d_i)$ increases the contribution of the divergence of the ion pressure tensor to the non-ideal out-of-plane electric field.}
\label{fig:heatflux}
\end{figure}

The importance of the ion scale physics is shown in Figs.~\ref{fig:heatflux}(a) and (b), in which the decomposition of the ion momentum equation at the time of maximum reconnection rate in a cut across the current sheet for $\lambda = 5 d_i$ is shown. As $k_i$ approaches $1/(3d_i)$, the role of the divergence of the ion pressure tensor in balancing the non-ideal electric field becomes more important over a broader region, whereas the contribution from ion inertia is significantly reduced, and the reconnection rate approaches the PIC rate. The broader diffusion region and reduction of the reconnection rate are consistent with the hybrid and PIC simulations of island coalescence \cite{stanier:2015}.

The scaling of the reconnection rate using $k_i = 1/(3d_i)$ is shown in Fig.~\ref{fig:gke10}. For smaller systems, the agreement is better than when $k_i = 1/d_e$ is used, but in the larger systems, the formation of plasmoids causes the maximum reconnection rates to level off in a similar manner, albeit at a lower absolute value, and reconnection proceeds to completion, unlike the PIC simulations.
 Thus, while these results show that the inclusion of the ion pressure tensor can reproduce some aspects of the PIC results for smaller systems, there are still discrepancies as larger systems are studied. Additionally, the current numerical model has $k$ as a free parameter for each species. Further work to improve the closure model or approximate $k$ self-consistently will be necessary \cite{wang:2015,dimits:2014}.

\section{Summary and conclusions}

We have performed a comparison between a variety of fluid models and PIC simulations \cite{karimabadi:2011} of how the island coalescence problem scales with system size. 

The resistive MHD simulations show flux pileup and sloshing, with an eventual transition from Sweet-Parker to plasmoid-dominated reconnection as the Lundquist number is increased, which is consistent with the literature \cite{knoll:2006,biskamp:1982}.

The Hall MHD simulations model large systems while maintaining $\delta \ll d_i$, causing X-point like current sheets to form and reconnection rates which show weak dependence on the system size. While the results are similar to Hall MHD studies of the Harris sheet \cite{birn:2001}, there are qualitative and quantitative differences when compared to the PIC simulations \cite{karimabadi:2011, stanier:2015}: The reconnection rates are 2-3 times larger and sloshing is not observed when reconnection remains in the Hall regime.

With the ten-moment two fluid model, which allows the electron and ion pressure tensors to depart from isotropy \cite{wang:2015}, we observe a constant maximum reconnection rate for larger systems, and a weak system-size dependence of the average reconnection rate, likely due to the secondary island formation. Though there is better agreement with the PIC results for small systems, the scaling of the reconnection rate is much weaker and reconnection proceeds to completion without sloshing, unlike the PIC case. A study on the effect of $k_i$ in the ion heat flux  shows that the non-isotropic ion pressure tensor is important in this geometry, in agreement with hybrid and PIC simulations \cite{stanier:2015}.

The results of this study highlight the difference between fluid and kinetic models of island coalescence. As the ratio of system size to ion skin-depth is increased, the reconnection rates in the Hall MHD and two fluid models decrease more slowly as compared to the PIC simulations, and there are qualitative differences such as the lack of sloshing in Hall MHD and secondary island formation in the two fluid model. While using a non-isotropic ion pressure tensor to capture the ion diffusion region in the two fluid model causes better agreement to be observed for small systems, there are still differences when the model is extended to larger systems. It follows that Hall MHD is insufficient to model the coalescence problem \cite{stanier:2015}, while further improvements need to be made to the heat flux closure used here to obtain better agreement with the PIC results in large systems.

\section{Acknowledgments}

This work is supported by NSF grant Nos. AGS-138944, AGS-1056898, AGS-14606169, DOE Contract DE-AC02-09CH11466, DOE Award DESC0006670 and NASA grant NNX13AK31G. This research used resources of the National Energy Research Scientific Computing Center, a DOE Office of Science User Facility supported by the Office of Science of the U.S. Department of Energy under Contract No. DE-AC02-05CH11231 and Trillian, a Cray XE6m-200 supercomputer at UNH supported by the NSF MRI program under grant PHY-1229408.

\bibstyle{plain}
\bibliography{reconnectionbib}

\end{document}